\documentclass[runningheads]{cl2emult}

\usepackage{makeidx}  
\usepackage{graphicx} 
\usepackage{subeqnar} 
\usepackage{multicol} 
\usepackage{cropmark} 
\usepackage{lnp}      
\makeindex            

\newcommand{\euler}[1]{{\usefont{U}{eur}{m}{n}#1}}

\newcommand{\umu}{\mbox{\euler{\char22}}}

\def\la{\mathrel{\mathchoice {\vcenter{\offinterlineskip\halign{\hfil
$\displaystyle##$\hfil\cr<\cr\sim\cr}}}
{\vcenter{\offinterlineskip\halign{\hfil$\textstyle##$\hfil\cr
<\cr\sim\cr}}}
{\vcenter{\offinterlineskip\halign{\hfil$\scriptstyle##$\hfil\cr
<\cr\sim\cr}}}
{\vcenter{\offinterlineskip\halign{\hfil$\scriptscriptstyle##$\hfil\cr
<\cr\sim\cr}}}}}

\def\ga{\mathrel{\mathchoice {\vcenter{\offinterlineskip\halign{\hfil
$\displaystyle##$\hfil\cr>\cr\sim\cr}}}
{\vcenter{\offinterlineskip\halign{\hfil$\textstyle##$\hfil\cr
>\cr\sim\cr}}}
{\vcenter{\offinterlineskip\halign{\hfil$\scriptstyle##$\hfil\cr
>\cr\sim\cr}}}
{\vcenter{\offinterlineskip\halign{\hfil$\scriptscriptstyle##$\hfil\cr
>\cr\sim\cr}}}}}

\begin{document}
\title*{Cosmic Infrared Background: ISOPHOT FIR source counts at 90, 150 and 180\,$\umu$m}
\toctitle{ISOPHOT CIRB project: FIR source counts 
\protect\newline at 90, 150 and 180\,$\umu$m}
%
%
\titlerunning{FIR source counts at 90, 150 and 180\,$\umu$m}
%
\author{Mika Juvela\inst{1}
\and Kalevi Mattila\inst{1}
\and Dietrich Lemke\inst{2}}
\authorrunning{Mika Juvela et al.}
%
%
\institute{Helsinki University Observatory, Helsinki, Finland
\and Max-Planck-Institut f\"ur Astronomie, Heidelberg, Germany}

\maketitle              

\begin{abstract}
Far-infrared maps obtained with ISOPHOT have been searched for point-like
sources. The majority of the 55 sources is believed to be extragalactic and
in most cases no previously known sources can be associated with them. Based
on the far-infrared spectral energy distributions it is likely that
dust-enshrouded, distant galaxies form a significant fraction of the
sources.

We estimate the number densities of extragalactic sources at 90$\umu$m,
150$\umu$m and 180$\umu$m wavelengths and at flux density levels down to
$\sim$100\,mJy. The counts are compared with models of galaxy evolution. The
counts exceed the predictions of current models, even those with strong
evolution, and no-evolution models are rejected at a high confidence level.

Comparison with recent results from COBE mission indicates that at 90$\umu$m
the detected sources correspond to $\ga$20\% of the extragalactic background
light. At longer wavelengths the corresponding fraction is 10\%.

\end{abstract}

\begin{table*}
\centering
\caption{The studied fields. The columns are:
name of the field, coordinates of the centre of each field, area of the map,
name of the ISOPHOT filter used in the observations, number of raster
positions observed, step between adjacent raster positions in
the staring mode mapping and integration time. The distance of adjacent
scans was in all cases identical to the raster step used along the scan
line}
\setlength\tabcolsep{5pt}
\begin{tabular}{lrrrrrrr}
\hline\noalign{\smallskip}
Field  &  \multicolumn{2}{c}{Map Centre} &  
                  Area & Filter  &  Rasters  & Step & $t_{\mathrm int}$  \\
      &    RA(2000.0) & DEC(2000.0)     & 
      (sq.degr.) &         &         &  (arcsec)  & (s)  \\
\noalign{\smallskip}
\hline
\noalign{\smallskip}
VCN    & 15 15 21.7 &  +56 28 58  &  0.030
                            & C\_90  &  10$\times$4  &  90 & 46  \\
       &            & & & C\_135 &  10$\times$4  &  90 & 46  \\
       &            & & & C\_180 &  10$\times$4  &  90 & 46  \\	

VCS    & 15 15 53.1 & +56 19 30   &  0.023
                            & C\_90  &  21$\times$2  &  90 & 46  \\
       &            & & & C\_135 &  21$\times$2  &  90 & 46  \\
       &            & & & C\_180 &  21$\times$2  &  90 & 46  \\	

NGPN   & 13 43 53.0 & +40 11 35   & 0.27
                            & C\_90  &  32$\times$4  &  180 & 23 \\
       &            & & & C\_135 &  32$\times$4  &  180 & 27 \\
       &            & & & C\_180 &  32$\times$4  &  180 & 27 \\
       & 13 42 32.0 & +40 29 06   & 0.53
                            & C\_180 &  15$\times$15 &  180 & 32 \\	

NGPS   & 13 49 43.7 & +39 07 30   & 0.27
                            & C\_90  &  32$\times$4  &  180 & 23 \\
       &            & & & C\_135 &  32$\times$4  &  180 & 27 \\
       &            & & & C\_180 &  32$\times$4  &  180 & 27 \\	

EBL22  & 02 26 34.5 & -25 53 43   & 0.19
                            & C\_90  &  32$\times$3  &  180  & 23 \\
       &            & & & C\_135 &  32$\times$3  &  180  & 27 \\
       &            & & & C\_180 &  32$\times$3  &  180  & 27 \\	

EBL26  & 01 18 14.5 & 01 56 40    & 0.27
                            & C\_90  &  32$\times$4  &  180  & 23 \\
       &            & & & C\_135 &  32$\times$4  &  180  & 23 \\
       &            & & & C\_180 &  32$\times$4  &  180  & 23 \\	

\hline
\end{tabular}
\label{table:maps}
\end{table*}

\section{Introduction}

In the far-infrared the cosmic infrared background (CIRB) consists of
radiation emitted by galaxies, intergalactic gas and dust, photon--photon
interactions ($\gamma$-ray vs. CMB) and, possibly, by decaying relic
particles. A large fraction of the energy released in the universe since the
recombination epoch is contained in the CIRB. An important aspect is the
balance between the UV-optical and the infrared backgrounds. With recent
observations at infrared and sub-mm wavelengths it has become obvious that
star formation efficiencies in the early universe derived from optical and
UV observations are underestimated (e.g. Madau et al. \cite{madau}; Steidel
et al.
\cite{steidel99}).

The analysis of the data from the COBE DIRBE (Hauser et al. \cite{hauser})
and FIRAS (Fixsen et al. \cite{fixsen98}) indicated a FIR CIRB flux at a
surprisingly high level of $\sim$1 MJy sr$^{-1}$ between 100 and 240
$\umu$m. Similar results had been obtained already by Puget et al.
\cite{puget96} and Schlegel et al. \cite{schlegel98}. Because of the great
importance of the FIR CIRB for cosmology these results require confirmation
by independent measurements.

The final goal of the ISOPHOT CIRB project is the determination of the FIR
CIRB flux level. First steps are the measurement of the CIRB fluctuations
and the detection of the bright end of FIR point source population. The
ISOPHOT CIRB project has potential advantages over the DIRBE analysis: (1)
with the much smaller f.o.v. ISOPHOT is capable of looking at the darkest
spots between the cirrus clouds; (2) in spite of the smaller f.o.v.
ISOPHOT's sensitivity surpasses that of DIRBE in the important FIR window at
120 -- 200 $\umu$m; (3) with the good spatial and spectral sampling the
galactic cirrus can be separated.

We have mapped four low-cirrus regions at high galactic latitude at the
wavelengths of 90, 150, and 180 $\umu$m. Here we report on the point sources
(galaxies) found in the FIR maps. The FIR source counts are important for
the study of the star formation history of the universe and for the testing
of models of galaxy evolution.

The point source extraction is based on the fitting of the detector
footprint to spatial data. The method is different from those used in most
previous studies (e.g. Kawara et al. \cite{kawara}; Puget et al.
\cite{puget99}) where the source detection has been based on the analysis
of the detector signal as a function of time. Our analysis is therefore
independent of and complementary to previous results.

\section{Observations}

The observations were performed with the ISOPHOT (Lemke at el.
\cite{lemke}) aboard ISO (Kessler et al. \cite{kessler}). The maps were
made in the PHT22 staring raster map mode (see Table~\ref{table:maps}). The
area covered is $\sim$1.5 square degrees. The fields have low surface
brightness and in some cases there is some redundancy i.e. the observed
pixel rasters partly overlap each other.

The data were processed with PIA (PHT Interactive Analysis) versions 7.1 and
7.2. The flux density calibration was made using the FCS (Fine Calibration
Source) before and after each map. The accuracy of the absolute calibration
is expected to be better than 30\% (Klaas et al. \cite{klaas98}).

Data reduction from the ERD (Edited Raw Data) to SCP (Signal per Chopper
Plateau) was performed using the pairwise method (Stickel \cite{stickel99}).
Instead of making linear fits to the ramps consisting of the detector
read-outs one examines the distribution of the differences between
consecutive read-outs. The mode of the distribution is estimated with myriad
technique (Kalluri \& Arce \cite{kalluri98}) and is used as the final signal
for each sky position. The pairwise method is robust against glitches and
the analysis was based on the data reduced with this method.

\section{Detection procedure} \label{sect:procedure}

The source detection was performed in two steps using data processed to the
AAP (Astrophysical Applications Data) level with PIA and the pairwise
method. The data consists of surface brightness values with error estimates.
Flat fielding was performed with custom routines.

Each surface brightness value was compared with the mean of the region
within a radius of $\sim$three times the size of the detector pixel. Values
more than 0.7$\sigma$ above the local background were considered as
potential point sources.
A model consisting of a point source and a constant background was fitted
into each region surrounding a candidate position. Footprint matrices were
used to calculate the contribution of the point source to the observed
surface brightness values. The free parameters of the fit were the source
flux density, the two coordinates of the source position, and the background
surface brightness. The formal errors are used to calculate the probability,
$P$, that the detection is not caused by background noise.

The completeness of the source detection and the number of false detections
were studied with simulations. The results were used to adjust the
probability level that was used for discarding uncertain detections. The
ratio $\rho$ between the source flux density and the background rms noise,
$\sigma_{\mathrm bg}$, was used as a criterion to discard uncertain sources.
Since $\rho$ is not directly related to the probability obtained from the
footprint fit it can be used as an additional safeguard against false
detections.

The actual source list consists of sources detected at two or three
wavelengths with a spatial distance between detections of less than
80\hbox{$^{\prime\prime}$}. The expected number of co-incidental
associations is no more than $\sim$10\% and does not significantly affect
the source counts.

\section{Source counts}

The surface density of sources is estimated by dividing the number of
detected sources at a given flux density level with the corresponding
`effective' map area.

As the first approximation the effective area corresponding to a given flux
density level was taken to be the sum of those maps where sources with equal
or lower flux densities were detected. 
The cumulative source densities obtained at 90$\umu$m, 150$\umu$m and
180$\umu$m are shown as histograms in Fig.~\ref{fig:allcounts1}. Two sets of
sources were used. The first set consists of all detections (dotted line)
while in the second set there are only sources detected at more than one
wavelength (solid line).
The results are similar for 150$\umu$m and 180$\umu$m and significant
differences are seen only at 90$\umu$m. This is expected, since sources seen
at 150$\umu$m are likely to be seen also at 180$\umu$m (and vice versa)
while more of the 90$\umu$m sources remain unconfirmed at the longer
wavelengths.

A third estimate for the cumulative source densities (dashed lines in
Fig.~\ref{fig:allcounts1}) was obtained by selecting sources based on the
ratio $\rho$ (see Sect.~\ref{sect:procedure}). For each flux density level
sources with $\rho>\rho_{\mathrm 0}$ were selected and no confirmation at
other wavelengths was required. The effective area was obtained by
integrating the total area where the rms noise was below 1/$\rho_{\mathrm
0}$ times the source flux density. The values of $\rho_{\mathrm 0}$ were
selected based on simulations. At the bright end the results agree with
earlier histograms since no sources are rejected and the corresponding areas
converge towards the total area mapped.

At very low flux density levels the small number of sources leads to large
uncertainties. The values obtained below 150\,mJy for 150$\umu$m and
180$\umu$m are probably only indicative.

\section{Discussion}

\subsection{Cirrus confusion}

We have checked the probability that some of the sources detected are small
scale cirrus structures (cirrus knots). 
Results of Herbstmeier et al. \cite{herbstmeier} show that at the scale of
the C100 beam size, $d\sim45$\hbox{$^{\prime\prime}$}, the expected cirrus
fluctuation amplitude is below 10\,Jy\,sr$^{-1/2}$ for all our 90$\umu$m
maps. According to Gautier et al. \cite{gautier92} this corresponds to a
flux density of 4\,mJy which is clearly below the flux densities of the
faintest 90\,$\umu$m detections. Therefore, cirrus is not likely to be a
significant contaminant in the source counts although, because of the
non-gaussian nature of the cirrus fluctuations (Gautier et al.
\cite{gautier92}), our source list may still contain a few cirrus
knots.

\subsection{Comparison with galaxy spectra} \label{sect:followup}

\begin{figure}
\centering
\includegraphics[width=.63\textwidth]{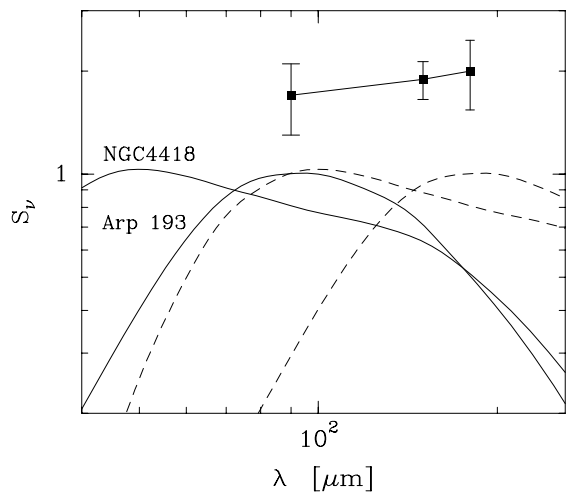}
\caption[]{
The average of the source spectra and the model spectra of Lisenfeld et al.
(\cite{lisenfeld}) for the luminous infrared galaxies Arp\,193 and
NGC\,4418. The dashed lines show the spectra of the two galaxies shifted to
$z$=1.0. The flux density scale is arbitrary}
\label{fig:lisenfeld}
\end{figure}

In Fig.~\ref{fig:lisenfeld} we compare the average of the source spectra
with the spectra of the galaxies Arp\,193 and NGC\,4418. In the sample of
luminous infrared galaxies presented by Lisenfeld, Isaak \& Hills
\cite{lisenfeld} Arp\,193 has the lowest and NGC\.4418 the highest
estimated dust temperature.

In the rest frame the SEDs of luminous infrared galaxies reach maxima
between 60$\mu$m and 100$\mu$m (Silva et al. \cite{silva}; Devriendt,
Guiderdoni \& Sadat \cite{devriendt}; Lisenfeld, Isaak \& Hills
\cite{lisenfeld}). The spectra of our FIR sources are flat in the observed
wavelength range and the emission peak is typically above 90$\mu$m. This is
consistent with most sources being at redshifts 0.5$\la$$z$$\la$1.

The FIR emission maximum of normal spiral galaxies is also located close to
100$\mu$m (e.g. Silva et al. \cite{silva}). However, in the case of a spiral
galaxy a FIR detection at the level of 0.1\,Jy would correspond to a visual
magnitude brighter than 16 and the optical counterpart should be readily
visible. The lack of clear visual counterparts indicates that most of our
sources are likely to be more distant and luminous infrared galaxies.

\subsection{Comparison with previous source counts}

From Fig.~\ref{fig:allcounts1} we obtain lower limits of
9$\times$10$^4$\,sr$^{-1}$, 2.0$\times$10$^5$\,sr$^{-1}$ and
2.2$\times$10$^5$\,sr$^{-1}$ for the cumulative source counts $\ge$100\,mJy
at 90\,$\umu$m, 150\,$\umu$m and 180\,$\umu$m, respectively. Similarly,
above 200\,mJy the source densities are 1.3$\times$10$^4$\,sr$^{-1}$,
6$\times$10$^4$\,sr$^{-1}$ and 7$\times$10$^4$\,sr$^{-1}$. The dotted
histogram, which lies mostly between the two other curves, gives at 100\,mJy
values 1.4$\times$10$^5$\,sr$^{-1}$, 2.5$\times$10$^5$\,sr$^{-1}$ and
3.5$\times$10$^5$\,sr$^{-1}$ at 90\,$\umu$m, 150\,$\umu$m and 180\,$\umu$m.

Kawara et al. \cite{kawara} have performed similar counts using ISOPHOT
observations of the Lockman Hole. The cumulative source counts were
1.1$\times$10$^5$\,sr$^{-1}$ and 1.3$\times$10$^5$\,sr$^{-1}$ at 95\,$\umu$m
and 175\,$\umu$m, respectively, for sources $S_{\nu}$$>$150\,mJy. At
90\,$\umu$m our results are close to their values. Our lowest estimates that
were based on detections at two or three wavelengths are, however, lower by
more than a factor of two. This could indicate that the requirement of
having detections at more than one wavelength does indeed underestimate the
counts.

Latest results from the ELAIS survey (Oliver et al.~\cite{oliver99}),
covering an area of 11.6 square degrees at 90$\umu$m, are similar to the
preliminary results reported by Rowan-Robinson et al. \cite{rr99}. At
$\sim$150\,mJy level the source density is 3.3$\times$10$^4$\,sr$^{-1}$
i.e. a factor of three lower than Kawara et al. \cite{kawara}. The result
is close to our lowest estimate as shown in Fig.~\ref{fig:allcounts1}.

Linden-V{\o}rnle et al. \cite{linden} have performed source counts in a 0.4
square degree field at 60$\umu$m and 90$\umu$m. At 90$\umu$m the
results are at 100\,mJy level similar to the ELAIS counts but are lower at
other flux density levels. Also in their re-analysis of the Lockman Hole
field they found significantly lower source densities than reported by
Kawara et al. \cite{kawara}.

Puget et al \cite{puget99} have published source counts based on 175$\umu$m
ISOPHOT observations of the Marano field covering $\sim$0.25 square degrees.
The source densities (e.g. $\sim$4$\cdot$10$^5$\,sr$^{-1}$ at 120\,mJy) are
slightly higher than our counts. 

The source counts obtained in these different ISOPHOT studies are mainly
within a factor of $\sim$2 from each other but direct comparison is made
difficult by the differences in the source flux calibration adopted
by the different authors. We have found a $\sim$30\% difference between the
DIRBE surface brightness values and our data calibrated with PIA version 7.3
(Juvela et al. \cite{vilspa}). This uncertainty is indicated in
Fig.~3 by the horizontal lines that are used to represent our source counts.
At 90$\umu$m the DIRBE calibration results in lower flux densities and at
150$\umu$m and 180$\umu$m in higher flux densities. These values are better
for the comparison with e.g. the results of Oliver et al.
\cite{oliver99} at 90$\umu$m since they used DIRBE as the basis of their
calibration.

\subsection{Comparison with galaxy models}

The source counts at 150$\umu$m and 180$\umu$m are much higher than
predicted by no-evolution models (e.g. Franceschini et al.
\cite{franceschini98}). At 180\,$\umu$m the difference is a factor of
five and these models can be safely rejected. Our results indicate that the
luminosity or the number of galaxies must evolve strongly with $z$.

Guiderdoni et al. \cite{guiderdoni98} have presented semi-analytic models
for the galaxy evolution. In their model E (which predicts the highest
source counts) both the `burst' mode of star formation rate and the relative
number of ultraluminous IR galaxies (ULIRGs) increase with $z$; at $z=5$
half of all galaxies are ULIRGs. The model is in good agreement with
extragalactic background light measurements in both optical and infrared.
The model predicts source counts of $\sim$1.7$\times$10$^5$\,sr$^{-1}$ and
$\sim$1.8$\times$10$^5$\,sr$^{-1}$ for sources brighter than 100\,mJy at
150\,$\umu$m and 180\,$\umu$m, respectively. The number of sources found in
this study clearly exceeds these predictions.

Franceschini et al. \cite{franceschini98} have presented similar models
which include contributions from two galaxy populations: dust-enshrouded
formation of early-type galaxies and late-type galaxies with enhanced
star-formation at lower redshifts. The predicted source counts at 170$\umu$m
are higher than in the model of Guiderdoni et al. \cite{guiderdoni98} and
thus in better agreement with our results.

\begin{figure}
\centering
\includegraphics[width=.43\textwidth]{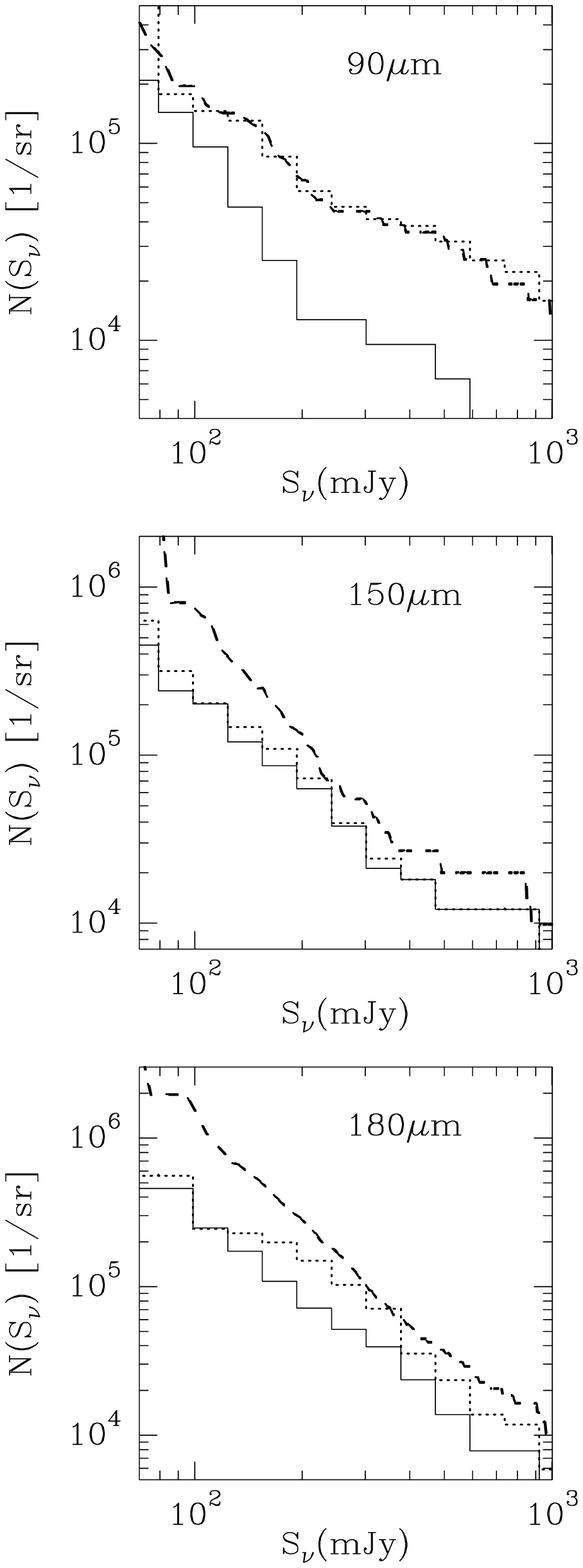}
\includegraphics[width=.55\textwidth]{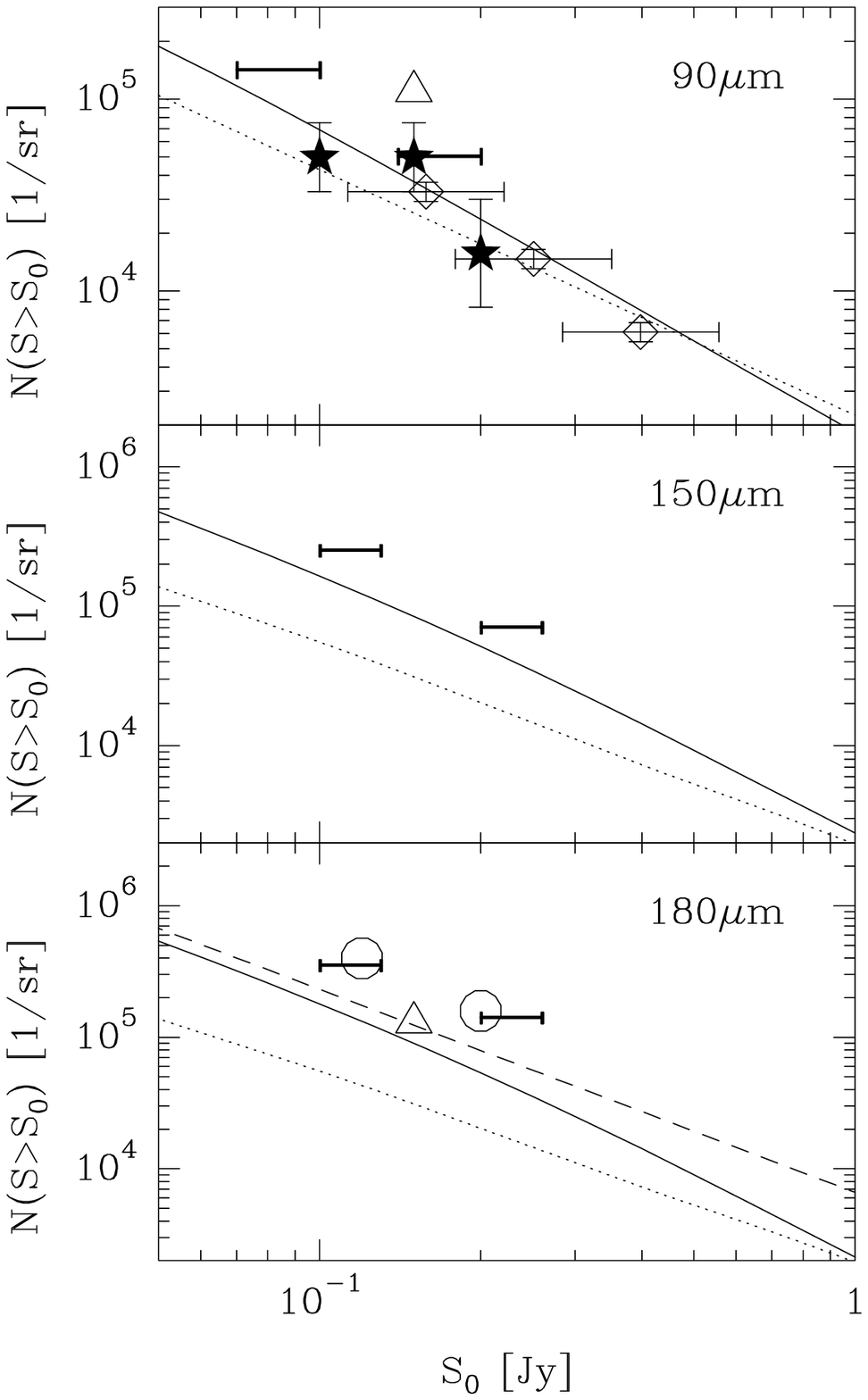}
\caption[]{
Cumulative source counts at 90$\umu$m, 150 $\umu$m and
180$\umu$m. The dotted line gives all sources detected and the solid line
sources detected at more than one wavelength. Areas were estimated according
to the faintest source detected in a map. The dashed curves represent cases
in which both the the source selection and the area determination were based
on the local background fluctuations (see text). \\
{\bf Fig. 3.} Comparison with other ISOPHOT counts and models of
galaxy evolution. Our source counts at 100\,mJy and 200\,mJy are shown
together with the results of Puget et al. \cite{puget99} (circles), Kawara
et al. \cite{kawara} (triangles), Oliver et al. \cite{oliver99}
(diamonds) and Linden-V{\o}rnle
\cite{linden} (stars).
We present our results as horizontal lines that indicate the difference
between the DIRBE calibration and the adopted ISOPHOT calibration (see
text). The predictions of model E of Guiderdoni et al. \cite{guiderdoni98}
are shown with solid lines and the evolutionary model of Franceschini et al.
\cite{franceschini98} with a dashed line.
Dotted lines show predictions of no-evolution models
(90$\umu$m: Guiderdoni et al. \cite{guiderdoni98}; 150$\umu$m and
180$\umu$m: Franceschini et al. \cite{franceschini98})}
\label{fig:allcounts1}
\end{figure}

In Fig.~3 we show the predictions of these two models
together with counts from other references. For this plot we have selected
from Fig.~\ref{fig:allcounts1} the values of the dotted histogram.

\clearpage
\addcontentsline{toc}{section}{Index}
\flushbottom
\printindex

\end{document}